\documentclass[prl,reprint,a4paper,superscriptaddress,preprintnumbers]{revtex4-2}
\usepackage{amsmath,amssymb,amsfonts,float,hyperref}

\def\bar{\overline}
\def\Re{\,\textbf{Re}}

\DeclareFontFamily{U}{BOONDOX-cal}{\skewchar\font=45 }
\DeclareFontShape{U}{BOONDOX-cal}{m}{n}{
	<-> s*[1.05] BOONDOX-r-cal}{}
\DeclareFontShape{U}{BOONDOX-cal}{b}{n}{
	<-> s*[1.05] BOONDOX-b-cal}{}
\DeclareMathAlphabet{\maths}{U}{BOONDOX-cal}{m}{n}
\SetMathAlphabet{\maths}{bold}{U}{BOONDOX-cal}{b}{n}
\DeclareMathAlphabet{\mathbs}{U}{BOONDOX-cal}{b}{n}

\begin{document}

\title{Standard Model from A Supergravity Model\\ with a Naturally Small Cosmological Constant}

\date{\today}

\author{Shing Yan \surname{Li}}
\email{sykobeli@mit.edu}
\affiliation{Center for Theoretical Physics, Department of Physics, Massachusetts Institute of Technology, Cambridge, MA 02139, USA}

\author{Yu-Cheng \surname{Qiu}}
\email{yqiuai@connect.ust.hk}
\affiliation{Department of Physics and Jockey Club Institute for Advanced Study, \\The Hong Kong University of Science and Technology, Hong Kong, S.A.R. China}

\author{S.-H. Henry \surname{Tye}}
\email{iastye@ust.hk, sht5@cornell.edu}
\affiliation{Department of Physics and Jockey Club Institute for Advanced Study, \\The Hong Kong University of Science and Technology, Hong Kong, S.A.R. China}
\affiliation{Department of Physics, Cornell University, Ithaca, NY 14853, USA}

\begin{abstract}
Guided by the naturalness criterion for an exponentially small cosmological constant, we present a string theory motivated 4-dimensional $\mathcal{N}=1$ non-linear supergravity model (or its linear version with a nilpotent superfield) with spontaneous supersymmetry breaking. The model encompasses the minimal supersymmetric standard model, the racetrack K\"ahler uplift, and the KKLT anti-$\rm D3$-branes, and use the nilpotent superfield to project out the undesirable interaction terms as well as the unwanted degrees of freedom to end up with the standard model (not the supersymmetric version) of strong and electroweak interactions.  
\end{abstract}

\preprint{MIT-CTP/5248}

\maketitle

{\it Introduction.}---The dark energy, or the observed cosmological constant $\Lambda_{\rm obs}$, poses a puzzle in fundamental physics. Its smallness with respect to the (reduced) Planck mass ($M_{\rm Pl} =2.4 \times 10^{18}$ GeV), $\Lambda_{\rm obs} \simeq 10^{-120} M_{\rm Pl}^4$, needs to be understood. Since $\Lambda$ is calculable in string theory, one is led to search for a reason within string theory. Hopefully, understanding its smallness will also reveal other structures in nature. Here is an attempt along this direction.
  
It has been shown that the string theory motivated racetrack K\"ahler uplift (RKU) model allows an exponentially small cosmological constant $\Lambda$ \cite{Sumitomo:2013vla}. Sweeping over a patch of the string landscape (i.e., scanning over the parameters of the model), one finds that the probability distribution $P(\Lambda) \propto \Lambda^{-1+k}$ for $\Lambda\to 0^+$, where $1 \gg k>0$. An exponentially small positive $\Lambda$ (e.g., the median value) naturally follows from the divergent behavior of such a properly normalized $P(\Lambda)$ at $\Lambda\to 0^+$. In this RKU model, the value of the observed $\Lambda_{\rm obs}$ implies the emergence of a new scale ${\bf m}$, which happens to be close to the  electroweak scale, ${\bf m}\sim m_{\rm EW} \sim 10^2\, {\rm GeV}$ \,\cite{Andriolo:2018dee}. The KKLT scenario  \cite{Kachru:2003aw} suggests that an anti-D3-($\bar{\rm D3}$)-brane can break supersymmetry (SUSY) and uplift the vacuum to a de Sitter (dS) vacuum. So one is led to include into the RKU model the contributions of the $\bar{\rm D3}$-brane tension $m_s^4$ as well as the effective Higgs potential $V_h$ after spontaneous symmetry breaking (SSB).  It is shown that their respective contributions must cancel to a high degree in order to maintain a naturally small $\Lambda$ in the RKU model \cite{Qiu:2020los},
\begin{equation}
	(\Delta V)_{\rm min} = \left|m_s^4+V_{h,{\rm min}}\right| < \Lambda \,,
	\label{eq:L}
\end{equation} 
where the negative $V_{h,{\rm min}} \simeq - \mathcal{O}(m_{\rm EW}^4)$. (In fact, there is no meta-stable dS vacuum solution in the RKU model if $\Delta V > \Lambda$.) This cancellation must be automatic; otherwise, fine-tuning is re-introduced. In this paper, let us focus on this condition, which is now a part of the naturalness criterion.  Since the Higgs potential  $|V_{h,{\rm min}}|$ is many orders of magnitude larger than $\Lambda_{\rm obs}$, one should expect $V_{h,{\rm min}}$ to be canceled or screened by some effect without fine-tuning  in any model with a naturally small $\Lambda$. In this paper,  we show how this is achieved in a specific SUGRA model with $\bar{\rm D3}$-brane(s).

Note that, in scanning over parameters of the model to obtain $P(\Lambda)$, we scan over all the parameters in $\Delta V$ \eqref{eq:L} too.  A naturally vanishingly small $(\Delta V)_{\rm min}$ with parameters spanning over (reasonable) ranges imposes tight constraints in model building. Not surprisingly, this has strong implications on particle physics phenomenology.

The motivation of the model comes from string theory.
In terms of the Planck mass scale, $m_s \simeq m_{\rm EW} \sim 10^{-16} M_{\rm Pl}$. In the brane world scenario with the warped flux compactification in Type IIB string theory [cf.\cite{Douglas:2006es,Becker:2007zj,Ibanez:2012zz,Baumann:2014nda}], this strongly suggests that both the standard model (SM) of strong and electroweak interactions and the $\bar{\rm D3}$-brane must sit at the bottom (or close to the bottom) of the same warped throat. This further suggests that the SM particles are simply open string modes inside the $\bar{\rm D3}$-brane (or a stack of $p\ge 5$ of them that span our 3-dimensional observable universe) \cite{Qiu:2020los}, or more generally, inside a $\bar{\rm D3}/{\rm D7}$ system, with a set of intersecting ${\rm D7}$-branes. Similar scenarios have been suggested earlier \cite{Cascales:2003wn,Antoniadis:2010hs,Kallosh:2015nia,GarciadelMoral:2017vnz,Cribiori:2019hod,Parameswaran:2020ukp}. In this paper, the naturalness criterion \eqref{eq:L} provides crucial guidance in developing the model. 

It was suspected that SUSY breaking by $\bar{\rm D3}$-brane(s) is spontaneous \cite{Kachru:2002gs,Kachru:2003sx,McGuirk:2012sb}. It turns out that this spontaneous SUSY breaking by an $\bar{\rm D3}$-brane can be incorporated into a non-linear SUSY model \cite{Volkov:1973ix}, which can be cast as a linear SUSY model with a nilpotent chiral superfield $X$, where $X^2=0$ \cite{Rocek:1978nb,Ivanov:1978mx,Ferrara:2014kva,Kallosh:2014wsa,Bergshoeff:2015jxa}. (This is analogous to a non-linear sigma model cast as a linear sigma model with a constraint.) To start the construction of the model,  we assume that the minimal supersymmetric standard model (MSSM) [cf.\cite{Martin:1997ns}] emerges from such a $\bar{\rm D3}/{\rm D7}$ system. 

Putting together the RKU model, the KKLT model and MSSM leads us to a 4-dimensional $\mathcal{N}=1$ supergravity model, namely a minimal non-linear (or nilpotent) SUSY standard model (mNSSM). In MSSM, some of the terms in the Higgs potential contribute an electroweak scale (semi-positive) vacuum energy density which is many orders of magnitude bigger than $\Lambda_{\rm obs}$. To satisfy \eqref{eq:L}, such terms have to be projected out, so the new resulting potential $\Delta V$ can vanish naturally after SSB. Our approach uses $X$ to impose constraints \cite{Lindstrom:1979kq,Komargodski:2009rz,DallAgata:2015zxp} on the particle spectrum as well as terms in the Higgs potential so that the model satisfies condition \eqref{eq:L}. 
As a result,  the Higgsinos and half of the two Higgs doublets are projected out, and the final Higgs potential for all practical purposes is precisely that in the SM, with the SM Higgs doublet $h=(h^0, h^-)$,
\begin{equation}
 \label{SMV}
	\Delta V =\left|m_s^2 - \kappa h^\dagger h\right|^2 \,,
\end{equation} 
where $(\Delta V)_{\rm min}=0$ after SSB, irrespective of the values of $m^2_s$ and the real positive $\kappa$. That is, the $\bar{\rm D3}$-brane tension $m_s^4$ and the Higgs potential $V_{h,{\rm min}}$ completely cancel each other to satisfy \eqref{eq:L}. 

Note that the supposed SUSY breaking from $\bar{\rm D3}$-brane(s) (in the KKLT scenario) is now completely screened by the Higgs potential. 
%As the $\bar{\rm D3}$-brane tension is canceled by the Higgs potential $V_{h,{\rm min}}$, 
As a consequence, the SUSY breaking in the model comes entirely from the K\"ahler uplift, which is exponentially small. This implies that a SUSY pair have almost degenerate masses; since superpartners have not been observed experimentally, all remaining superpartners in MSSM must be removed, again employing $X$ to project them out. The final model is for all practical purposes precisely the SM of strong and electroweak interactions. However, the superpartners of the (closed) string modes (such as the graviton) that are not confined to the 
$\bar{\rm D3}$-brane(s) are expected to remain.\\

{\it Model.}---In the Wilsonian effective field theory approach, we write down a low energy $4$-dimensional $\mathcal{N}=1$ supergravity model with parameters calibrated at the electroweak energy scale. Besides the contribution of the $SU(3)\times SU(2) \times U(1)$ gauge supermultiplets, the model is specified by the K\"ahler potential $K$ and the superpotential $W$. (Bar on quantities implies complex conjugate.)
\begin{widetext}
\begin{align}
\label{mNSSM}
	K&=-2\ln{\left[\left(T+\bar{T}-X\bar{X}-n_u H_u^\dagger H_u -n_d H_d^\dagger H_d +K_{\rm matter}\right)^{3/2}+\frac{\xi}{2}\right]}\;,\nonumber\\
	W&=W_0(U_i,S) +W_{\rm np}(T)+ \tilde{\mu} H_u H_d - X\left(\tilde{m}_s^2+\tilde{\gamma} H_u H_d\right) +W_{\rm matter}\;, \nonumber\\
	&\quad W_{\rm np}(T)=Ae^{-aT} +Be^{-bT},  \qquad \xi=-\frac{\zeta(3)}{4\sqrt{2}(2\pi)^3}\chi(\mathcal{M})\left(S+\bar{S}\right)^{3/2}>0\; .
\end{align}
\end{widetext}
Here, $K_{\rm matter}$ and $W_{\rm matter}$ contain the quark and lepton superfield contributions in MSSM, including all the Yukawa couplings in $W_{\rm matter}$. The complex structure moduli $U_i$ and the dilaton $S$ are stabilized \cite{Giddings:2001yu,Kachru:2003aw,Qiu:2020los}, thus $W_0$ can be treated as a constant, leaving only the K\"ahler modulus $T$ which measures the volume of the flux-compactified Calabi-Yau orientifold $\mathcal{M}$. The K\"ahler uplift is provided by $\xi$, a stringy $\alpha'^3$-correction \cite{Becker:2002nn}; with multiple $U_i$, the Euler index of the manifold $\chi(\mathcal{M})<0$, so $\xi>0$. We expect $\xi \sim \mathcal{O}(10^{-2})$ in the KU model \cite{Balasubramanian:2004uy,Westphal:2006tn,Rummel:2011cd,deAlwis:2011dp,Sumitomo:2012vx,Louis:2012nb}. The non-perturbative $W_{\rm np}$, which can come from gaugino condensation in $D7$-branes, provides the racetrack. Together, they form the RKU model when $T$ is being stabilized. In the brane world picture in string theory, $U_i$, $S$, $T$ and the graviton-gravitino pair are close string modes while the rest are open string modes inside the $\bar{\rm D3}$-brane(s) that span our observable universe. 

Here the two Higgs doublets $H_u$ and $H_d$ in MSSM are explicitly displayed, since only Higgs fields acquire vacuum expectation values (vev) in SM and contribute to $\Lambda$. We denote $H_u H_d=(H_u)_i(H_d)_j\epsilon^{ij}=H_u^+H_d^--H_u^0H_d^0$. $n_u$ and $n_d$ are $\mathcal{O} (1)$ normalization constants. $\tilde{\gamma}$ describes the (dimensionless) coupling between $X$ and Higgs superfields. The scalar potential $V$ is given by two parts, namely the $F$-terms and the $D$-terms.
% which could be rearranged into several contributions. 
Possible $D$-term contribution from the geometrical sector is ignored here.  In units where $M_{\rm Pl}=1$,
\begin{align}
	V&=K_{I\bar{J}} F^I \bar{F}^{\bar{J}}-3 e^K W\bar{W}+V_D\nonumber\\
	&=V_T+V_X+V_{H,F}+V_{H,D}=V_T + \Delta V\;,
	\label{TV}
\end{align}
where $F^I=e^{K/2}K^{I\bar{J}}D_{\bar{J}}\bar{W}$, $D_I=\partial_I+K_I$, $K_I=\partial_IK$ and $K^{I\bar{J}}K_{I\bar{J}}=1$.
%Repeated notation $I$ indicates all degrees of freedom.
\begin{subequations}
\label{FDterms}
\begin{align}
	V_T&=e^K K^{T\bar{T}}\left|D_TW\right|^2-3e^K \left|W\right|^2\;,\label{eq:vt}\\
	V_X&=K_{X\bar{X}}F^X\bar{F}^{\bar{X}}+\big(K_{T\bar{X}}F^T\bar{F}^{\bar{X}}+{\rm c.c.}\big)\;,\label{eq:vx}\\
	V_{H,F}&=K_{H\bar{H}}F^{H}\bar{F}^{\bar{H}}+\big(K_{H\bar{I}}F^H\bar{F}^{\bar{I}}+{\rm c.c.}\big)\;,\label{eq:vhf}\\
	V_{H,D}&=\sum_a \frac{1}{2}g_a^2 D^{a 2}\;.\label{eq:vhd}
\end{align}
\end{subequations}
 In the RKU model where only $V_T$ is present, it has a no-scale structure when $\xi=0$ and a small positive $\Lambda$ emerges after $T$ is stabilized with $\xi$ turned on. $g_a$ is gauge coupling and $D^a$ is the corresponding Killing potential in supergravity language. Contributions \eqref{eq:vx},\eqref{eq:vhf} and \eqref{eq:vhd} are all semi-positive squares. To minimize $V$ (to satisfy condition \eqref{eq:L}), we need $\Delta V=V_X+V_{H,F}+V_{H,D} < \Lambda$, that is, for all practical purposes, each term should vanish by itself. To achieve that, we introduce several algebraic constraints \cite{Komargodski:2009rz,DallAgata:2015zxp,DallAgata:2016syy}
 to project out some undesired degrees of freedom as well as interactions; explicitly,
\begin{subequations}
\label{allp}
\begin{align}
		X^2&=0\;,\label{proj0}\\
	X\bar{H}&={\rm chiral}\;, \label{proj1} \\
		X\left[\left(H_u\right)_i-\epsilon_{ij} \left(\bar{H}_d\right)^j\right]&=0\;,\label{proj2} \\
XQ_i=XL_j=XW_{\alpha} &=0\;.\label{susyout} 
\end{align}
\end{subequations}
In short, \eqref{mNSSM} together with \eqref{allp} defines our mNSSM model. 
Here the constraint \eqref{proj1} (i.e., $\bar{\maths{D}}_{\dot{\alpha}}\left[X \bar{H}_i\right]=0$) is on every Higgs chiral superfield, and $\epsilon_{ij}\epsilon^{jk}=\delta_i^k$ in \eqref{proj2}. In \eqref{susyout}, index $i$ runs over all quark chiral superfields, the index $j$ runs over all lepton chiral superfields and index $\alpha$ runs over all the field strength chiral superfields of the vector superfields. 
In the context of string theory ($\bar{\rm D3}$-brane in an orientifold), constraint \eqref{proj0} is discussed in Ref.\cite{Kallosh:2014wsa,Bergshoeff:2015jxa}, while constraints \eqref{proj1} and \eqref{susyout} are discussed in Ref.\cite{Vercnocke:2016fbt,Kallosh:2016aep}.

As we shall see,  $V_X=0$ is automatic after SSB, while constraint \eqref{proj1} eliminates the term $V_{H,F}$ and the constraint \eqref{proj2} eliminates $V_{H,D}$ terms (note that both $V_{H,F}$ and $V_{H,D}$ are present in MSSM). The last set \eqref{susyout} of constraints project out the remaining R-parity odd degrees of freedom, i.e., the scalar quarks and scalar leptons as well as the gauginos in MSSM;  as we shall see, this  is necessary since SUSY breaking is negligibly small in the model. \\

{\it Properties.} ---The nilpotent superfield $X$ emerges from the presence of an $\bar{\rm D3}$-brane \cite{Ferrara:2014kva,Kallosh:2014wsa,Cribiori:2019hod}. The constraint $X^2=0$ \eqref{proj0} projects out the scalar degree of freedom of $X$. Expanded in supergravity variables $\Theta$ \cite{Wess:1992cp},
\begin{equation}
	X= \maths{x}+\sqrt{2}\Theta G+\Theta^2 F^X \quad{\rm and }\quad \maths{x}={GG}/{2F^X}\;,
\end{equation}
where $G_\alpha$ is the goldstino. Since the expectation value $GG=0$, $\maths{x}$ and any field component in \eqref{FDterms} that contains $G_\alpha$ (due to constraints \eqref{allp}) will drop out in $V$. Therefore, only the $K_{X {\bar X}}$ term remains in \eqref{eq:vx} and we have
\begin{equation}
	V_X=\frac{\left|\tilde{m}_s^2+\tilde{\gamma} H_u H_d\right|^2}{3\left(T+\bar{T} + \cdot \cdot \cdot \right)^2}\;.
	\label{fx}
\end{equation}
Next, let us consider a quark chiral superfield $Q= \maths{q}+\sqrt{2}\Theta \psi+\Theta^2 F^Q$. The constraint $XQ=0$ \eqref{susyout} yields \cite{Komargodski:2009rz}
\begin{align}
Q&=\frac{\psi G}{F^X} -\frac{G^2 F^Q}{2\left(F^X\right)^2} +\sqrt{2}\Theta \psi+\Theta^2 F^Q\nonumber\\
&=\sqrt{2}\left(\psi-\frac{F^QG}{F^X}\right)\tilde{\Theta} +F^Q \tilde{\Theta}^2 \;,\label{XQi}
\end{align}
where $\tilde{\Theta} = \Theta + G/\sqrt{2}F^X$.
Here the scalar quark is projected out.  The other constraints are similar.

{\it Higgs Sector.}---$V_{H,F}$ \eqref{eq:vhf} in $V$ \eqref{TV} will in general violate \eqref{eq:L} without fine-tuning, so we like to remove it. Constraint \eqref{proj1} on a Higgs superfield yields, in global SUSY \cite{Komargodski:2009rz},
\begin{align}
H &=\maths{h} +\sqrt{2} \theta \Psi_H + \theta^2 F^H \,, \\
\Psi_H&=i \sigma^{\nu}\left(\frac{\bar G}{{\bar F}^{\bar X}}\right) \partial_{\nu}\maths{h}\, ,\nonumber \\
F^H&= -\partial_{\mu}\left(\frac{\bar G}{{\bar F}^{\bar{X}}}\right){\bar \sigma}^{\nu} \sigma^{\mu} \frac{\bar G}{{\bar F}^{\bar{X}}} \partial_{\nu} \maths{h} +\frac{1}{2\big({\bar F}^{\bar{X}}\big)^2} \bar{G}^2 \partial^2\maths{h}\,, \nonumber
\label{eq:6b}
\end{align}
where the corresponding expressions in SUGRA can be found in Ref.\cite{DallAgata:2015zxp}. We see that, not only $\Psi_H$ is removed,
so is the corresponding auxiliary field $F^H$. Applying the constraint \eqref{proj1} on every Higgs superfield, \eqref{eq:vhf} yields, after setting $G_{\alpha}=0 $ and so $F^H=0$,
\begin{equation}
\label{mu2term}
	V_{H,F}\propto |{\tilde{\mu}}|^2\left(H_u^{\dagger}H_u+H_d^{\dagger}H_d\right)+\cdots  \;\to\;0 \, .
\end{equation}
This means that $\tilde{\mu} H_u H_d$ in $W$ \eqref{mNSSM} contributes to the scalar potential (of order $\mathcal{O} (m_{\rm EW}^6/M_{\rm Pl}^2)$) only in $V_T$, which is  already included in the RKU model \cite{Qiu:2020los}. 

Now let us expand the whole scalar potential and focus on Higgs fields. Relevant and marginal terms are kept and higher order terms which is suppressed by $M_{\rm Pl}$ are ignored. Having in mind that $T+\bar{T}\sim\mathcal{O}(10^3)$ (so the string scale is close to the grand unified scale), we only keep terms proportional to $(T+\bar{T})^{-2}$. Due to the no-scale structure of the K\"ahler potential, Higgs fields receive field normalization according to their respective kinetic terms (after the $T$ stabilization),
\begin{equation}
	h_u=\maths{h}_{u}\left(\frac{3n_{u}}{T+\bar{T}}\right)^{\frac{1}{2}}\;, \quad
	h_d=\maths{h}_{d}\left(\frac{3n_{d}}{T+\bar{T}}\right)^{\frac{1}{2}}\;,
	\label{eq:nolh}
\end{equation}
so that kinetic terms of $h_u$ and $h_d$ are in canonical form, which means that they are the observed Higgs fields in MSSM. 
%Here $\maths{h}$ stands for scalar component of superfield $H$. 
The remaining dependence of $(T+\bar{T})$ in the potential can be absorbed into the parameters, 
\begin{equation}
	m_s=\tilde{m}_s\left[3\left(T+\bar{T}\right)^2\right]^{-\frac{1}{4}}\;,\quad
	\gamma=\tilde{\gamma}\left(27n_u n_d\right)^{-\frac{1}{2}}\;,
\end{equation}
where the $D$-term potential is the same as that in MSSM. There should be some $T$-dependence in $V_{H,D}$, which could be further absorbed into $n_u$ and $n_d$ after normalization of Higgs fields. We write down the potential for two Higgs doublets with all parameters rescaled to observed value properly,
\begin{align}
	V_{2h}&=2 m_s^2\Re\left(\gamma h_u h_d\right)+ \left|\gamma h_u h_d\right|^2\nonumber\\
	+&\frac{g^2+g'^2}{8}\left(\frac{\left|h_u\right|^2}{n_u}-\frac{\left|h_d\right|^2}{n_d}\right)^2+\frac{g^2}{2}\left|\frac{h_u^\dagger h_d}{\sqrt{n_u n_d}}\right|^2\;,
	\label{eq:2h}
\end{align}
where the first two terms come from $V_X$ and the last two terms from $V_{H,D}$; $g$ and $g'$ are the $SU(2)$ and $U(1)$ gauge couplings respectively, and $h_u=(h_u^+, h_u^0)$, $h_d=(h_d^0, h_d^-)$ as in MSSM. With the $SU(2)$ symmetry, we choose SSB as $\langle h_u^0  \rangle$ develops a vev, followed by $\langle h_d^0  \rangle \ne 0$. Parameter $\gamma$ is a complex number in principle. It 
can be made real by rotating the complex fields $h_u^0$ and $h_d^0$, so that $\gamma >0$ and  $\gamma \langle h_uh_d \rangle = - \gamma   \langle h_u^0 \rangle   \langle h_d^0 \rangle  = - \gamma v_uv_d <0$.

After SSB, $v_uv_d =m_s^2/\gamma$, $v_d^2= m_s^2\sqrt{n_d/n_u}/\gamma $, so $v_u^2=m_s^2 \sqrt{n_u/n_d}/\gamma$.  Following MSSM, we introduce $\tan \beta=v_u/v_d$. It turns out that, at the minimum,
\begin{equation}
 \tan\beta =\sqrt{\frac{n_u}{n_d}}\to
\begin{cases}
 	V_{H,D, {\rm min}}=0\\
 	V_{2h, {\rm min}}=-m_s^4 
\end{cases} \to V_{X, \rm min}=0\;,
\label{Vmm}
\end{equation} 
so $\Delta V=0$ and condition \eqref{eq:L} is satisfied. Notice that the result $\Delta V=0$ is insensitive to the values of $m_s, \gamma, g^2, g'^2, n_u, n_d$ and $\mu$. 

If the $|\tilde{\mu}|^2$ term in $V_{H,F}$ \eqref{mu2term} is present, as in MSSM, then $V_{2h}$ \eqref{eq:2h} will have an additional term (setting $n_u=n_d=1$ and with a rescaled $\mu$): $V_{2h} \to V_{2h} + |\mu|^2(|h_u|^2 + |h_d|^2)$. In this case  $V_{2h, {\rm min}} = -(m_s^2-|\mu|^2/\gamma)^2$ yielding $\Delta V= m_s^4 -(m_s^2-|\mu|^2/\gamma)^2 >0$, so condition \eqref{eq:L} is not satisfied except via fine-tuning. This is the reason we impose \eqref{proj1} to obtain \eqref{mu2term}; as a consequence, Higgsinos are absent in this model.

Although  the above result \eqref{Vmm} with $V_{H,D, {\rm min}}=0$ is desirable, it is straightforward to check that the charged Higgs boson mass is the same as the $W$-boson mass, $m_{H^\pm}^2=m_W^2=(v_u^2+v_d^2)g^2/2$, which is disastrous, as experiments show such $H^{\pm}$ do not exist.
Fortunately, this problem can be easily solved by the constraint \eqref{proj2}, which projects out $V_{H,D}$ \eqref{eq:vhd} as well as half of the Higgs degrees of freedom. 

Let us explain the choice of the particular linear combination in the constraint \eqref{proj2}. Similar to \eqref{XQi}, a linear combination of the Higgs scalar fields is projected out. Suppose, instead of \eqref{proj2}, we impose
$X[(H_u)_i - q \epsilon_{ij} (\bar{H}_d)^j]=0$, where $q$ is real but arbitrary. This projection gives a linear relation between observed (properly normalized according to \eqref{eq:nolh}) Higgs fields,
\begin{equation}
\label{6c}
h_u^+=-q\sqrt{\frac{n_u}{n_d}}\bar{h}_d^-\;,\quad h_u^0=q\sqrt{\frac{n_u}{n_d}}\bar{h}_d^0\;.
\end{equation}
Now $h_u^\dagger h_d=\bar{h}_u^+ h_d^0+\bar{h}_u^0 h_d^-=q\sqrt{n_u/n_d}(-h_d^-h_d^0+h_d^-h_d^0)=0$, $h_uh_d=-q\sqrt{n_u/n_d}h_d^\dagger h_d\equiv -\sqrt{n_d/n_u} h^\dagger h/q$ and the two Higgs doublet potential $V_{2h}$ \eqref{eq:2h} reduces to the one Higgs doublet potential\;,
\begin{align}
	V_h=&-2m_s^2 \frac{\gamma}{q}\sqrt{\frac{n_d}{n_u}} h^\dagger h+\frac{\gamma^2}{q^2}\frac{n_d}{n_u} \left(h^\dagger h\right)^2 \nonumber\\
	&\qquad+\frac{g^2+g'^2}{8}\frac{\left(q^2-1\right)^2}{q^4n_u^2}\left(h^\dagger h\right)^2\, ,
	\label{2hVD}
\end{align}
where the last term in $V_{2h}$ \eqref{eq:2h} has dropped out. In this case, $V_{D,H} > 0$, which in general violates condition \eqref{eq:L} unless $q=1$. After SSB, instead of \eqref{Vmm},  one is sitting at 
\begin{align}
	V_{h,\rm min}^{(q)}=-\frac{m_s^4}{1+\cfrac{g^2+g'^2}{8\gamma^2q^2n_un_d}\left(q^2-1\right)^2}\;,
\end{align}
implying $\Delta V >0$ and condition (1) is not satisfied unless $q=1$, when  
 $V_{h,\rm min}^{(q)}$ is lowest. That is, constraint \eqref{proj2} is energetically preferred, i.e., dynamically determined.

With constraint \eqref{proj2}, or $q=1$ for \eqref{6c}, the $V_{H,D}$ term in \eqref{2hVD} drops out, and the Higgs potential for a single Higgs doublet becomes
\begin{equation}
	V_h=-2\kappa m_s^2 h^\dagger h + \kappa^2 \left(h^\dagger h\right)^2 \,,
\end{equation}
where $\kappa =\gamma\sqrt{n_d/n_u}$ is taken to be a real positive (dimensionless) coupling. Since the constraint \eqref{proj2} removes $V_{H,D}$ in \eqref{eq:2h}, the gauge couplings do not enter $V_h$. This is exactly the SM Higgs potential and its contribution comes entirely from $V_X$,	$V_X=m_s^4+V_h=|m_s^2 - \kappa h^\dagger h|^2$, as shown in \eqref{SMV}.
 Note that the quartic coupling $\kappa^2 > 0$.

After SSB, the Higgs field acquires a vev,
	$\langle h^\dagger h\rangle={v^2}/{2}={m_s^2}/{\kappa}$.
Expanding around this minimum, physical massive Higgs boson and three massless Goldstone bosons emerge,
\begin{equation}
	m_h^2=4m_s^2\kappa \;,\quad m_{G^0}^2=m_{G^\pm}^2=0\;.
\end{equation}
Putting in the observed $m_H=125$ GeV and $v=246$ GeV, one finds $\kappa=0.36$ and $m_s=104.3$ GeV. The triple and quartic Higgs self-couplings are the same as those in SM. The potential value for $V_X$ \eqref{SMV} after SSB is 
	$V_{X,{\rm min}}= 0$, irrespective of the values of $m_s^2$ and the real positive $\kappa$; with  
	$$\Delta V= V_{F,H}+V_{H,D}+V_X=V_{X}\;,$$
we now have $(\Delta V)_{\rm min} = V_{X,{\rm min}}=0$,
satisfying the condition \eqref{eq:L} without fine-tuning. One may choose to extend the model so 
$$\Delta V=\left| m_s^2 - \kappa h^\dagger h + \delta\left(h^\dagger h\right)^2 \right|^2\, ,$$
where the real coefficient $\delta \sim \mathcal{O}(1/M_{\rm Pl}^2)$. We see that $(\Delta V)_{\rm min}=0$ is robust.

Note that the SUSY breaking $\bar{\rm D3}$-brane tension is completely screened by the Higgs potential, so $\bar{\rm D3}$-brane(s) do not break SUSY in this model. SUSY breaking and the uplift from AdS to dS vacuum comes from the $\xi$ term only. However, the SUSY breaking by the K\"ahler uplift is very small ($m_{3/2}^2 \sim \mathcal{O} (\Lambda/M_{\rm Pl}^2)$), so any SUSY pair will have almost degenerate masses, which is a problem as no superpartner has been observed. Again, this phenomenological problem can be easily solved by imposing the constraints \eqref{susyout}, which removes all the scalar quarks and scalar leptons as well as the gauginos. Constraint \eqref{proj1} projects out the Higgsinos while constraint \eqref{proj2} reduces the two Higgs doublets to a single doublet, and implies the same Yukawa couplings as those in SM. \\

{\it Summary.}---The mNSSM model is given by \eqref{mNSSM} together with \eqref{allp}. After imposing the constraints \eqref{allp}, the particle spectrum of the mNSSM model is reduced to exactly that in the SM (in energy scales around or below $m_{\rm EW}$), except for  the very light modes of $S$, $U_i$ and $T$ (and the graviton-gravitino pair). Since these modes are not confined to the $\bar{\rm D3}$-brane(s), their superpartners (dilatino, the Uenos and the Tino) cannot be projected out by $X$. Here, the role of $\bar{\rm D3}$-brane(s) is not to break SUSY (as opposed to the KKLT model \cite{Kachru:2003aw}), but to cancel the electroweak Higgs potential contribution to $\Lambda$.

Once the contributions to $\Lambda$ from the $\bar{\rm D3}$-brane tension cancels the Higgs potential $V_{h,\rm min}$, the value of $\Lambda$ follows from the $T$ stabilization in the RKU model \cite{Sumitomo:2013vla,Qiu:2020los}. There, the divergent behavior of the probability distribution $P(\Lambda) \sim \Lambda^{-1+k}$ around $\Lambda \sim 0^+$ follows when we scan over the ratio $z=A/B$ of the coefficients of the two non-perturbative terms in $W_{\rm np}=Ae^{-aT} +Be^{-bT}$ \eqref{mNSSM}, while the value of $1 \gg k >0$ depends on the ratio $\beta =b/a$ (as we scan over $a$ and $b$). 
Here the gravitino gains a tiny mass from the $\xi$-induced SUSY breaking, which is for the whole orientifold, not limited to the $\bar{\rm D3}$-branes or the throat we live in. \\

 {\it Comments.}---Some comments are in order here.
 
$\bullet$ Both the Higgs potential $V_{h, \rm min} = - m^4_{\rm EW}$ and the (necessary in MSSM) SUSY breaking scale $m_s^4 \gtrsim m^4_{\rm EW}$ are many orders of magnitude bigger than the observed $\Lambda$. Each by itself will generate a huge contribution to $\Lambda$ in the absence of fine-tuning. Therefore, it is neat to arrange them to precisely cancel each other without fine-tuning, a necessary condition for a naturally small $\Lambda$ in mNSSM. 
%This leads to strong implications on particle physics phenomenology. The model illustrates that the naturalness criterion can be a powerful tool in exploring fundamental physics. 
We believe that this or a similar cancellation is necessary in any model for a naturally small $\Lambda$. 

$\bullet$ In fitting experimental data, the couplings and masses are taken to have physical (i.e., renormalized) values, as measured at the electroweak energy scale, \`a la the Wilsonian approach.
 It is important to study the radiative corrections of the model. It is argued that the statistical approach in the RKU model does not have the usual radiative instability problem \cite{Tye:2016jzi}. Intuitively, this should not be too surprising. To determine $P(\Lambda)$, one scans over ranges of all the parameters of the model that yield a local minimum solution; typical radiative corrections contribute tiny shifts when compared to the variations of the parameters that are already included in the scanning.% However, if radiative correction shifts $\Delta V$ so that $\Delta V > \Lambda$, there is no minimum solution \cite{Qiu:2020los}, and ranges of parameters leading to this situation should not be included in the scanning that determines of $P(\Lambda)$.

 $\bullet$ The dS vacuum solution with a small $\Lambda$ in the RKU model is only meta-stable. Fortunately, tunneling to the true vacuum will take much longer than the age of our universe \cite{Tye:2016jzi}.
 
$\bullet$ Looking back at the KPV picture \cite{Kachru:2002gs}, where a small stack of $\bar{\rm D3}$-branes sits in a ``false" vacuum, we see that the electroweak contribution, by lowering the vacuum energy of the ``false" vacuum, tends to suppress the tunneling probability when compared to that in the KPV scenario.

$\bullet$ In the model, one expects Kaluza-Klein modes as well as string excitation modes to be present at scales not too far from $m_{\rm EW}$. Their detection will provide strong evidence for the scenario. Furthermore, their spectra will reveal valuable details of the warped throat and exactly where and how the SM sits in the throat. 

$\bullet$ The mNSSM has no particle that is a suitable candidate for cold dark matter. However, it has very light scalar fields like the $U_i$,  $S$ and  $T$ \cite{Rummel:2011cd,Sumitomo:2012vx,Tye:2016jzi}, one (or some) of which can play the role of super-light bosonic particles in the fuzzy dark matter scenario \cite{Hu:2000ke,Marsh:2015xka}.  

$\bullet$ The model can easily accommodate the $D3-{\bar{\rm D3}}$ inflationary model, \`a la the KKLMMT scenario \cite{Kachru:2003sx}. This scenario should happen in the early universe in another (less warped) throat in the orientifold. 

$\bullet$ In the model, we introduce degrees of freedom and terms in the Higgs potential to be projected out later. It will be interesting to find a more direct formulation without going through this seemingly tortuous path. 

%$\bullet$ One can choose to incorporate some of the higher order terms: e.g., $X(m_s^2+ \gamma H_uH_d)$ in $W$ can be extended to $X(m_s^2+ \gamma H_uH_d + \sigma (H_uH_d)^2)$ and ${\tilde{\mu}} H_uH_d \to {\tilde{\mu}} H_uH_d + \tau (H_uH_d)^2$. It will be interesting to explore their properties.

$\bullet$ We hope the results here, may be together with the speculation on the quark and lepton mass distributions \cite{Andriolo:2019gcb}, provide hints in the search for the SM in Type IIB string theory and/or F theory.

$\bullet$ With the emergence of a scale ${\bf m}\sim 10^2$ GeV,  the electroweak scale is natural in the RKU model \cite{Andriolo:2018dee}, so we do not need the ``technically natural" SUSY phenomenology. In the present model, motivated by superstring theory, SUSY is underpinning the model  but it is not quite detectable experimentally. However, we like to point out that it is possible, with some modification of the mNSSM, to bring back some of the superpartners where they are heavy. As an illustration, we may replace $X(m_s^2 +\gamma H_uH_d)$ in $W$ by $X(M_s^2e^{-2cT} +\gamma H_uH_d)$. In this case, after $T$ stabilization, we acquire a new scale $M_s=m_se^{+cT}$, where we can choose $M_s \sim \mathcal{O} (10^{10})$ GeV, and use it to generate soft terms to raise the masses of some superpartners so that they can now remain in the particle spectrum.\\

We thank Cliff Burgess, Keith Dienes, Renata Kallosh, Ling Feng Li, Tao Liu, Liam McAllister, Fernando Quevedo, Gary Shiu and Timm Wrase for valuable communications and  comments. 
This work is supported in part by the AOE grant AoE/P-404/18-6 issued by the Research Grants Council (RGC) of the Government of the Hong Kong SAR China.

\bibliography{ref}
\bibliographystyle{apsrev}

\end{document}